\documentstyle{ARTICLE}
\begin{document}

Differing Interactions Require Baryon and Lepton Conservation

R. Mirman


\newpage

Baryon and lepton numbers are conserved. Why? Baryon number must be because
baryons are subject to strong interactions, leptons are not. Conservation
of baryons leads to that of leptons. This raises further questions which
are noted.

\newpage

A fundamental attribute of elementary particles is their set of conserva-
tion laws. Some arise from the nature of space [ref. 1, p.27, 299], but
others are nongeometrical, such as conservation of baryons, and following
from it, conservation of leptons. Conservation of baryon, so of lepton,
numbers, comes not from space but from their interactions.

This discussion is schematic, excluding everything unneeded, which
simplifies the writing and aids understanding, and by not including the
irrelevant, emphasizes the generality of results. We ignore electromag-
netism and gravitation and their labels. They do not allow decay of
protons. To be concrete, and as it is often discussed this way, we speak of
the decay of the proton, but this applies to any baryon (a fermion with
strong interactions), and antibaryon, and pions, but that can be any meson
(a boson with strong interactions). Also lepton (a fermion with no strong
interactions) means either lepton or antilepton. All three types of
particles have weak interactions, that would cause, if it were possible,
the proton to decay into leptons plus pions, or directly into leptons (it
does not matter if the decay is attributed to other interactions; the label
weak merely denotes the interaction, whatever it is, whatever its
properties, that causes the decay that we wish to show is impossible).
Strong interactions are often stated as plural, for generality. As there
are no other stable particles (known), the analysis indicates that if there
are several types, every baryon is affected by each.

Why is baryon number conserved? If it were not a particle with strong
interactions would go into ones without; strong interactions would be
"turned off". But weak interactions cannot "turn off" strong ones. There is
no Hamiltonian with one interaction that changes another interaction. We
first give an intuitive picture, then a formal argument. Purely
heuristically, to see what goes wrong, take the unphysical case of the
proton having weak interactions, the pion not. This gives a Feynman diagram
in which the proton emits a virtual pion and then decays into three lep-
tons. But the pion, being virtual, has to be reabsorbed. The lep- tons,
however, do not interact with it, so it cannot be reabsorbed leaving it in
a very unpleasant situation --- implying that protons cannot so decay.
Similarly, a state only of leptons, with no strong interactions, cannot go
into one with baryons (without an equal number of antibaryons). The reac-
tion is not possible in either direction.

While the weak interaction of baryons cannot turn the strong one off or on,
mesons do go into leptons. Why? The photon is analogous; its number is not
conserved, though charge is. However it is neutral, it couples to a neutral
object, taken as a particle-antiparticle pair (the current is of this
form). Electron-photon scattering can be regarded as the creation of a pair
by the photon; the positron annihilates the electron, and the electron of
the pair replaces it. So while the electromagnetic interaction (of an elec-
tron) cannot be altered, photons can be taken as not having a direct inter-
action. Their creation or annihilation does not modify an interaction.
Similarly for mesons, which can be deemed to have no strong interactions,
but rather couple to particle-antiparticle pairs. We can view meson-baryon
scattering as annihilation of the baryon by the antiparticle of the pair,
and its replacement by the baryon of the pair. Because of what they are
coupled to, mesons can decay into objects not affected by strong interac-
tions.

With this intuitive understanding of why baryons cannot decay into only
leptons, but mesons can, we turn to a formal analysis. To study the decay
we consider the action of the Hamiltonian on a proton. Acting on a state at
t = 0, time-translation operator exp(iHt), H is the Hamiltonian,
gives the state at time t. Can this take a baryon to a state whose
fermions are only those not having strong interactions? The effect of this
is seen from that of H, a sum of the free particle Hamiltonians for the
proton, pion and leptons, plus interaction terms, for the weak and strong
interactions. The state of the system is a sum of terms, one the state of
the proton, another (if decay were possible) the product of pion and lepton
states (plus products of lepton states if there were those decays), each
sums over other labels and integrals over momenta or space. Take the ini-
tial state as a proton, say at rest. The free part of H changes its
phase. The weak interaction part, were decay possible, decreases its coef-
ficient in the sum, while increasing that of the (say) pion plus lepton,
initially zero --- starting as a proton, the state becomes a sum of the
proton, its contribution decreasing, plus the pion plus lepton state, with
increasing contribution. For the decaying pion the behavior is similar:
starting as a pure pion it becomes a sum of that plus a state of leptons,
with the contribution of the first decreasing, of the second increasing.

What is a proton? We define it as a particle obeying Dirac's equation, with
mass m(P), where this equation includes the (irrelevant and suppressed)
electromagnetic, weak and strong interactions (whose forms are not needed).
It is the presence of these interactions that determines what a proton is.
(Correctly a particle is an eigenstate of the two Poincar\'e invariants
[ref. 2, p.114]. For a free particle, and one with an electromagnetic
interaction, Dirac's equation is equivalent. Whether this is true with weak
and strong interactions seems unknown so consequences of, perhaps
important, differences, if any, are not clear. And it is possible that put-
ting interactions in invariants, which must be done whether Dirac's equa-
tion is used or not, might limit them. Particles are also eigenstates, or
sums, of the momentum operators [ref. 3, p.93], of which the Hamiltonian is
one. We ignore these, and refer to Dirac's equation as it is familiar, but
the discussion could be of invariants, which might be revealing). The
statefunction of the proton then is a solution to coupled nonlinear equa-
tions. We need information about it but cannot solve explicitly so
represent it in a way that allows analysis, using an expansion. The argu-
ments though are exact; we do not calculate, so need not truncate.

The physical particle, labeled with a capital, that obeying Dirac's equa-
tion with all interactions, is a sum of states (schematically):
summing over all states to which the proton is
connected by interactions, including any number of pions, and so on. The
summations represent ones over internal labels and integrals over momenta.
These, and all other irrelevancies, like spin, are suppressed. That they
can be shows the generality of this. State  p) is the function
satisfying Dirac's equation with the weak interaction, but not the strong
--- its effect is given by this sum, which is thus an eigenstate, with mass
m(P), of the total Hamiltonian, including all interactions. Individual
terms in the sum differ in energy and momenta; it is the sum that has the
eigenvalues. We can take pi) as either bare or physical; for the
former each term is an infinite sum, which is encapsulated by regarding it
as physical.

The coefficients are determined by the requirement that this be a solution
of the complete Dirac's equation, and normalization (P P) = 1. Also
the initial state, say a proton at rest, gives c(x,0), a wavepacket, with
all other c's = 0, at t = 0; they depend on the statefunction for
 P). These particles are virtual in not obeying the physical Dirac's
equation, that with interactions. Only P is physical. Dirac's equation
(with interactions schematic, and higher-order terms not excluded) is

States of different numbers of pions are orthogonal, so this gives
an infinite set of coupled equations for the c's;
solving (in principle) gives the state of the (physical) proton. (We
need not consider how far this can be taken to find, and solve, recursion
relations to obtain physical states; doing so for a particle that does
decay might provide information about it, and its other interactions.) This
expansion has an infinite number of pions, that is terms p)pi)**r,
for all r geq 0, because (writing  p) and  pi)
using creation and annihilation operators acting on the vacuum) the strong
part of the interaction Hamiltonian is (schematically)
is the annihilation, a* the creation, operators for p), the
b's are the ones for pions. This acting on  p)pi )**r gives
p)pi)**(r+1) and p)pi)**(r-1), so the expan                              -
sion includes all r geq  0.

Now the weak interaction acts on  p) supposedly causing it to decay,
so the final state is

showing the transition to, say, a pion plus a lepton, and to
three leptons, with coefficients of non-occurring terms zero. The energy of
 fs) is m(P), not m(p), so needs contributions from sum
c(p)pi) l)pi )pi ), and so on. However  fs)
is, say, a lepton plus a pion, so these other terms, to which this is
orthogonal, cannot contribute. The decay of the proton cannot conserve
energy, thus cannot occur. Likewise decays of leptons (the tau) to
baryons are similarly ruled out.

Contrast this with the strong decay of the Delta; the physical one is
Delta(p), while Delta satisfies Dirac's equation without the strong
interaction (thus with no interaction); pi(p)  is the physical pion,
pi satisfies interaction-free equations. Then Delta(p)  has an expan-
sion similar to the proton's, except that it can decay by the strong inter-
action into a proton plus a pion giving extra terms (with coefficients w),
           The Delta ), having no interac-
tions, does not decay, so this expansion cannot be used for an argument
like that for the proton's weak decay ---  p) and  Delta )
are solutions of the free-particle Dirac's equation, there is only a single
interaction the strong one (which causes the decay), so the argument fails.
This decay, of the entire sum, is possible. (But Delta ) cannot
decay into particles with no strong interaction.)

The neutron has an actual weak decay; why do the arguments not apply?
Taking it at t = 0 in a single particle state, gives
obeying Dirac's equation with the weak interaction, but not the strong, and
 N) is the physical particle, with both interactions. The n)
decays to p) l) l)', so this becomes
The d-
coefficients give the physical state  P) l) l)'. The state
then is
the first coefficient decreases in
time, the second increases, so the total probability is constant. These two
(orthogonal) states have equal energy. The neutron can decay into a
strongly-interacting particle (the proton), but not otherwise, because it
is state  n) that decays to  p). But the physical neutron i              s
a sum of terms like n)pi )**r, so the resultant proton is a
sum of such terms as p)pi)**r, which is the physical proton
state --- the proton, affected by the strong interaction, is such a sum
(plus others appearing also for the neutron). If the neutron were to decay
into state L) without the strong interaction, the expansion would
have terms like L)pi )**r, which do not sum to a state of any
physical particle --- there is no final state with this expansion so the
matrix element of the Hamiltonian causing such decay is zero; decays
annulling the strong interaction can- not occur.

Conservation of baryons implies that of leptons. The numbers of fermions in
all states must all be odd, or all even, and we assume that the number of
particles minus antiparticles is constant (it is not clear that otherwise a
hermitian Hamiltonian is possible that would not lead to difficulties like
those above, especially if neutrinos have nonzero mass, as we expect [ref.
3, p.70]). Then leptons cannot be produced or destroyed if baryon (minus
antibaryon) number is constant --- lepton number is con- served.

The pion can decay into leptons: Why does the argument not apply? The
physical pion pi(p)) can be written
obeys Dirac's equation with strong interactions, but not the weak --- this
causes the decay --- and the prime indicates antiparticles. This contains
two terms (plus irrelevant ones for other particles coupled to the pion).
The strong part of the pion's Hamiltonian
acting on a pion gives a particle-
antiparticle pair, acting on this pair gives the pion, and similarly for
r pions which is mixed with the pair plus r-1 pions. Only two states
appear in the expansion for the pion, unlike the infinite number for the
proton. The weak part of H acting on  pp') causes it to decay to
two leptons; this is related by crossing to proton-antiproton scattering
into two leptons, and also to the decay n Rightarrow p + 2l, which we
saw is possible. Thus both terms in the sum for the physical particle con-
tribute to the decay, it does conserve energy (and momentum), therefore
cannot be ruled out.

The argument for electric-charge conservation is the same. Charge conserva-
tion is related to gauge invariance, a partial statement of Poincar\'e
invariance [ref. 3, p.43] --- this relates an allowed interaction to the
Poincar\'e group. An interaction violating charge conservation would not
transform under gauge transformations as other terms in the Hamiltonian,
giving Poincar\'e transformations (on massive objects) that induce gauge
transformations (on massless ones) resulting in physically-identical obser-
vers who undergo the different gauge transforma- tions --- these cannot be
fully specified --- thus physically identical, but who see different
Hamiltonians. The Hamiltonian would not be well-defined, implying
inconsistent physics. It is fortunate that charge is conserved.

There are other implications requiring investigation; we mention them in
hope of stimulating such.

All interactions known are of lowest order. Why? For electromagnetism
linearity is enforced by gauge (Poincar\'e) invariance [ref. 3, p.57]. For
strong interactions, take a particle, a Delta or p, that emits a pion.
Higher-order terms would couple it not to a single pion, but to more.
Intuitively we can guess why only lowest order occurs since it gives
diagrams which we interpret (purely heuristically) as two, or more, pions
emitted sequentially. Higher-order means that these are emitted together.
However this is the limit of the lowest order in which the time between
emissions goes to zero. A higher-order interaction would be this limit,
which is included in the lowest order as one case; higher-order terms
adding nothing, would be irrelevant. Summing all diagrams, and integrating
over time, would give contributions from terms that have the same effect as
higher-order ones, thus changing only the value of the sum, so the value of
the coupling constant --- an experimental parameter (at present), thus we
could not distinguish contributions from terms of different order, implying
higher order would be undetectable. This regards particles as virtual. But
consider a decay in two steps, each emitting a pion. If the intermediate
object's life were sufficiently short, this would be equivalent to pions
being emitted simultaneously. If a nucleon had an interaction of the form
NNN' pi, the emission of an NN' pair could be thought of as due to the
decay of a pion, and the interaction taken as the limit of the emission of
a pion, and then its decay, when its lifetime becomes zero, merely changing
the sum.

Suppose that the only particles were nucleons and pions, no excited ones.
With a linear interaction final states of pion-nucleon scattering have only
a single pion --- the nucleon could not store the extra energy and decay to
a second pion. Nonlinear interactions give states with more than one pion,
but could be simulated by short-lived excited states --- the pion scatters
and excites the nucleon which then decays giving a second pion. Nonlinear
interactions would be indistinguishable from existence of excited states;
perhaps we could require all interactions be linear with addition of
excited states. Why is the gravitational interaction of lowest order? The
interaction of matter with gravity is clear [ref. 3, p.73]. But what
determines how matter fixes the gravitational field [ref. 3, p.150]? The
tensor could be T(mu nu )T(nu rho ), or could it? These are some
questions that should be looked at.

Gravitation raises another point. There are reasons why it may be unable to
act on scalar particles: pions, kaons, ... [ref. 3, p.61]. Assuming that it
acts on vectors, like rho, then objects affected by gravity decay to
ones that are not (the rho to pions), and ones not affected go to ones
that are (the pion to leptons). Does the argument not forbid this? It fails
because there is no such thing as a state of, say, two gravitons [ref. 3,
p.187], nor even a free graviton, since gravity is --- necessarily --- non-
linear [ref. 3, p.69]. The question remains open. Except for a few
particles, it is not even known (experimentally) which have interactions
with gravity, nor what these are --- something of great interest. Theoreti-
cal prejudice should not substitute for analysis and experiment. This
emphasizes the difference between gravity and other interactions, and the
value of these questions as a probe into the laws of nature.

There are strong constraints on physics as seen in many ways [ref. 1, p.2;
ref. 2; ref. 3]. It is well known that interactions are greatly limited by
the properties of space, but also because of other interactions. Perhaps
this analysis, and more the questions it leads to, can induce further
inquiry. There are reasons for the laws of nature, and it is fortunate that
reasons, and laws, are such that we are able to find, and understand, them.

1. R. Mirman, {\em Group Theory: An Intuitive Approach} (Singapore: World
Scientific Publishing Co., 1995).

2. R. Mirman, {\em Group Theoretical Foundations of Quantum Mechanics}
(Commack, NY: Nova Science Publishers, Inc., 1995).

3. R. Mirman, {\em Massless Representations of the Poincar\'e Group, elec-
tromagnetism, gravitation, quantum mechanics, geometry} (Commack, NY: Nova
Science Publishers, Inc., 1995).

\end{document}